\documentclass[twocolumn,amsmath,amssymb,pre]{revtex4}
\usepackage{graphicx,color,longtable}

\begin{document}

\author{Alex Malins}
\affiliation{School of Chemistry, University of Bristol, Cantock Close, Bristol, BS8 1TS, UK}
\affiliation{Bristol Centre for Complexity Science, University of Bristol, Bristol, BS8 1TS, UK.}

\author{Jens Eggers}
\affiliation{School of Mathematics, University Walk, Bristol BS8 1TW, UK.}

\author{Hajime Tanaka}
\affiliation{Institute for Industrial Science, The University of Tokyo, Komaba-4-6-1, Meguro-ku, Tokyo, 153-8505, Japan.}

\author{C. Patrick Royall}
\affiliation{HH Wills Physics Laboratory, Tyndall Avenue, Bristol, BS8 1TL, UK}
\affiliation{School of Chemistry, University of Bristol, Cantock Close, Bristol, BS8 1TS, UK}
\affiliation{Centre for Nanoscience and Quantum Information, Tyndall Avenue, Bristol, BS8 1FD, UK}

\email{paddy.royall@bristol.ac.uk}

\title{Lifetimes and Lengthscales of Structural Motifs in a Model Glassformer}

\begin{abstract}
We use a newly-developed method to identify local structural motifs in a popular model glassformer, the Kob-Andersen binary Lennard-Jones mixture. By measuring the lifetimes of a zoo of clusters, we find that 11-membered bicapped square antiprisms, denoted as 11A, have longer lifetimes on average than other structures considered. Other long-lived clusters are similar in structure to the 11A cluster.
These clusters group into ramified networks that are correlated with slow particles and act to retard the motion of neighbouring particles. The structural lengthscale associated with these networks does not grow as fast as the dynamical lengthscale $\xi_4$ as the system is cooled, in the range of temperatures our molecular dynamics simulations access. Thus we find a strong, but indirect, correlation between static structural ordering and slow dynamics. 
\end{abstract}

\maketitle

\section{Introduction}

The nature of the rapid increase in viscosity as liquids are cooled toward the glass transition is the subject of many theoretical approaches, however there is no consensus on its fundamental mechanism~\cite{cavagna2009,berthier2011,stillinger2013,biroli2013}. One plausible scenario is the emergence of self-induced memory effects upon supercooling of liquids, which causes slow dynamics~\cite{gotze2008}. However the recent discovery of dynamic heterogeneities, i.e., spatial heterogeneities in the relaxation dynamics that emerge on supercooling \cite{hurley1995,yamamoto1997,ediger2000,berthier2011e}, is suggestive of the importance of a growing dynamic length scale in the slowing down approaching the glass transition. 

In addition to these dynamical phenomenona, the idea of a structural change leading to vitrification has a long history~\cite{frank1952}. Sir Charles Frank suggested that upon supercooling liquids would form polyhedral motifs such as icosahedra that do not fill space. A related approach, geometric frustration~\cite{tarjus2005}, suggests that the glass transition can be thought of as a manifestation of a crystallisation-like transition that would occur in curved space (where the structural motifs tessellate~\cite{nelson2002}), but that growth of the ``crystal nuclei'' of polyhedral motifs is frustrated in Euclidean space. Another approach based on frustration against crystallization (due to random disorder effects or competing orderings)~\cite{tanakaGJPCM,tanaka2012} addresses the origin of slow dynamics and the physical factors controlling glass-forming ability within the same framework.

It has become clear that a range of glass formers exhibit a change in structure upon the emergence of slow dynamics~\cite{shintani2006,coslovich2007a}, and it is debated as to whether there exists a static lengthscale that underlies the growing lengthscale for the dynamical correlations. 
Broadly speaking two types of structure have been identified: spatially extendable crystal-like ordering ~\cite{shintani2006,tanaka2010,sausset2010,leocmach2012}, and non-extendable polyhedral ordering ~\cite{jonsson1988,kondo1991,tomida1995,jullien1996,dzugutov2002,lerner2009,pedersen2010,coslovich2011,malins2013jcp}. For the former it has been suggested that critical-like fluctuations of crystalline order are the origin of dynamic heterogeneities in certain classes of supercooled liquid~\cite{tanaka2010}. The latter concerns particles organised into polyhedra that cannot tile Euclidean space due to geometrical constraints~\cite{frank1952,tarjus2005}. Instead they form ramified structures with a fractal dimension that is less that the dimensionality of the system, i.e. non-extendable ordering. Some metallic glasses have been shown to exhibit this second type of ordering~\cite{schenk2002,miracle2004}. The exact relationship of the polyhedral order to the dynamic heterogeneities is unclear, however measurements have shown that the polyhedral domains are slow to relax~\cite{coslovich2007a}. 
Although the two types of orderings have a different nature, we note that both are induced to lower the free energy locally in the situation where its global minimization (crystallization) is prohibited~\cite{tanaka2012}. In relation to this, we note that even polyhedral ordering often has some connection to crystalline order, e.g., for example icosahedra in quasicrystal approximants~\cite{pedersen2010}. Furthermore, even in the case 
where extendable order dominates slow dynamics, non-extendable polyhedral order competes with extendable crystal-like order in systems 
such as 2D spin liquids~\cite{shintani2006} and polydisperse hard spheres~\cite{leocmach2012}.

One of the main difficulties with identifying structural correlations in supercooled liquids is that it is not known \textit{a priori} which (if any) type of static order is important for the dynamic slowdown. Frequently studies have employed structure detection methods, such as Voronoi face analysis~\cite{tanemura1977}, common neighbour analysis~\cite{honeycutt1987}, bond orientational order analysis~\cite{steinhardt1983}, or the topological cluster classification employed here~\cite{williams2007,royall2008,malins2013tcc} that search for predefined types of structural ``motif'' and see how the numbers change as a function of temperature~\cite{dellaValle1994,jonsson1988,tomida1995,royall2008,malins2013jcp}. However recently a number of ``order-agnostic'' schemes have been devised to identify structural correlations without first having to define what structures will be searched for~\cite{biroli2008,mosayebi2010,sausset2011,kob2011non,cammarota2012a,hocky2012,dunleavy2012}. In relation to this, it has recently been pointed out that for accessing such hidden structural ordering in apparently random structures it is crucial to focus on many-body correlations~\cite{leocmach2013,coslovich2013}.

To strengthen the link between structure and glassy behaviour three types of evidence have been presented. Firstly, many-body structural correlation functions have been presented that clearly show structural changes occur on supercooling towards the glass transition~\cite{coslovich2007a,pedersen2010,coslovich2013}. Secondly, dynamically slow regions have been correlated with different types of local ordering~\cite{dzugutov2002,coslovich2007a}. Thirdly, the presence of structural and dynamic length scales that grow similarly has been sought~\cite{shintani2006,tanaka2010,sausset2010}. This is motivated by strong evidence that, for sufficient cooling (which leads to relaxation timescales that may or may not be  accessible to computer simulations as we employ here), an increasing dynamical lengthscale necessitates an increasing lower bound to a structural lengthscale~\cite{montanari2006}. The case of growing structural and dynamic length scales is controversial. One of us has identified a direct correspondence between the growing dynamical and structural lengthscales in polydisperse systems displaying crystal-like ordering~\cite{tanaka2010,sausset2010,leocmach2013}, while others have claimed that structural lengthscales are decorrelated from the dynamic lengthscales and only grow weakly on cooling~\cite{charbonneau2012,charbonneau2013,charbonneau2013a}.

Local structural ordering has yet to be found in some glassforming systems. For these systems no one-to-one correspondence between an order-agnostic structural lengthscale and the dynamical lengthscale has been found~\cite{widmerCooper2006,kob2011non,charbonneau2012,dunleavy2012}, but static perturbation analysis has suggested a growing structural lengthscale~\cite{mosayebi2012}. 

The Kob-Andersen binary Lennard-Jones system~\cite{kob1995a} of interest here is known to display polyhedral ordering~\cite{coslovich2007a}. In this system, weakly growing structural lengthscales have been identified by the order-agnostic ``point-to-set'' analysis~\cite{hocky2012}, while other approaches using static perturbation of inherent structures~\cite{mosayebi2010} and finite size scaling~\cite{karmakar2012b} find a stronger increase in static lengthscales. The polyhedral ordering in the Kob-Andersen mixture~\cite{kob1995a}, take the form of the bicapped square antiprism [``11A'' in the topological cluster classification (TCC) nomenclature~\cite{williams2007,royall2008,malins2013tcc,malinsThesis} owing to its original identification as minimum energy cluster of the Morse potential~\cite{doye1995,doye1997}]. This structure
has been linked to slow dynamics~\cite{coslovich2007a} and frustrates crystallisation, which in the KA mixture occurs spontaneously by phase separation into two face-centred cubic lattices~\cite{toxvaerd2007}. We note that crystals based on 11A could in principle coexist with other structures [the stoichimetry of the 11A crystal with a small particle in the centre (see Section \ref{sectionClusterComposition} and Fig. \ref{figGComp}) is not compatible with the KA mixture~\cite{fernandez2003}]. Another possibility is four-fold symmetric crystals which have been predicted as low-lying energy minima for the KA mixture~\cite{middleton2001}. An order parameter associated with the formation of 11A clusters has been shown to control the dynamical phase transition in trajectory-space, a hallmark of dynamical facilitation theory for the glass transition~\cite{chandler2010}, indicating that the transition has both structural and dynamical character~\cite{speck2012}. Furthermore, ultra-stable glasses of the KA mixture have been found to be rich in 11A ~\cite{singh2013}. Here we study the spatial correlations between the domains of 11A in the supercooled liquid.

We consider the lifetimes of a multitude of structures in the supercooled liquid using the TCC algorithm. We detail our simulation protocol in section \ref{sectionSimulationDetails}, and briefly review the KA model's dynamical behaviour in section \ref{sectionDynamicalBehaviour}.
We use the TCC to identify any structural changes that occur in these mixtures on cooling towards the glass transition in section \ref{sectionStructuralAnalysis}. We then study in section \ref{sectionClusterLifetimeDistributions} the lifetimes of the clusters that are found at deeply supercooled state points in order to gauge which structures are likely candidates to be associated with slow domains of dynamic heterogeneities. Finally we analyse how correlation lengths for the domains of structured particles are related to the growing dynamic lengthscale in section \ref{sectionCorrelationLengths} before concluding.

\section{Model and simulation details}
\label{sectionSimulationDetails}

The Kob-Andersen (KA) binary mixture is composed of 80\% large (A) and 20\% small (B) particles of the same mass $m$~\cite{kob1995a}. The nonadditive Lennard-Jones interactions between each species, and the cross interaction, are given by $\sigma_{AA}=\sigma$, $\sigma_{AB}=0.8\sigma$, $\sigma_{BB}=0.88\sigma$, $\epsilon_{AA}=\epsilon$, $\epsilon_{AB}=1.5\epsilon$, and $\epsilon_{BB}=0.5\epsilon$. The results are quoted in reduced units with respect to the A particles, i.e. we measure length in units of $\sigma$, energy in units of $\epsilon$, time in units of $\sqrt{m\sigma^2/\epsilon}$, and set Boltzmann's constant $k_\mathrm{B}$ to unity. 
The interactions are truncated and smoothed using the Stoddard-Ford method~\cite{stoddard1973}. The truncation lengths are in proportion to the interaction lengths~\cite{kob1995a}, i.e. $r_\mathrm{tr}^{AA}=2.5$, $r_\mathrm{tr}^{AB}=2.0$ and $r_\mathrm{tr}^{BB}=2.2$. The simulations consist of $N=10976$ particles in 3D with periodic boundary conditions such that $N_A=8781$ and density $\rho=1.2$. The $\alpha$-relaxation time $\tau_{\alpha}^{A}$ for each state point is defined by fitting the Kohlrausch-Williams-Watts stretched exponential to the decay of the intermediate scattering function (ISF) of the $A$-type particles. 

Equilibrated samples at each temperature were prepared by simulating for $100\tau_{\alpha}^{A}$ in the canonical $NVT$-ensemble using the Nos\'{e}-Poincar\'{e} thermostat with coupling parameter $1.0$~\cite{nose2001}. The thermostat was switched off, then further equilibration was performed in the microcanonical $NVE$-ensemble using the velocity Verlet algorithm for $1000\tau_{\alpha}^{A}$. On completion of the equilibration process, trajectories of length $300\tau_{\alpha}^{A}$ were sampled for analysis. Colder state points were obtained in a step-wise fashion by quenching instantaneously from an equilibrated configuration of the previous higher temperature state point. The stability of the deep quenches was checked by ensuring there was no time evolution in the ISF, the partial radial distribution functions $g_{AA}(r)$, $g_{AB}(r)$ and $g_{BB}(r)$, or the number of clusters detected by the TCC algorithm across the trajectories. Crystallisation was not seen in any of the simulations.

\section{Dynamical Behaviour}
\label{sectionDynamicalBehaviour}

\begin{figure}
\begin{centering}
\includegraphics[width=7cm]{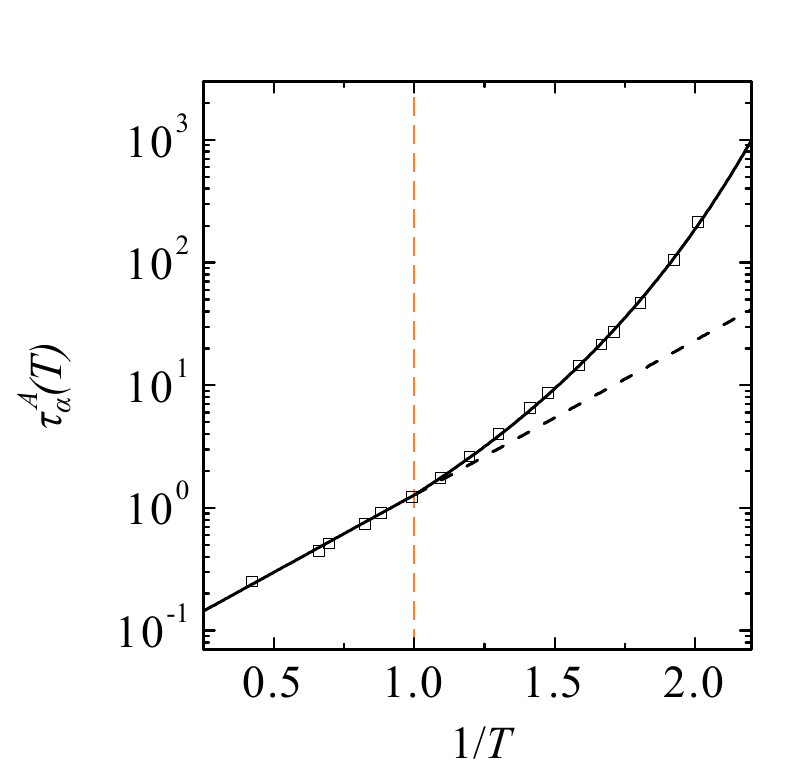}
\par\end{centering}
 
\caption{Increase in the alpha relaxation time, $\tau_\alpha^A$, on cooling. The data are fitted with a hybrid Arrhenius-VFT  fit (solid lines). The dotted lines indicate the relaxation times predicted by the high-$T$ Arrhenius fit. Orange dashed line indicates the crossover from Arrhenius to super-Arrhenius ($T^\ast$).
\label{figAngell}}
\end{figure}

In Fig. \ref{figAngell} the relaxation time $\tau_\alpha^A$ as a function of inverse temperature is plotted. For the equilibrium liquid state points at high temperatures, the relaxation times are well fitted by an Arrhenius function. As the temperature is lowered a cross-over in the dynamical behaviour occurs and the relaxation times increase faster than predicted by the Arrhenius equation~\cite{goldstein1969,berthier2003,berthier2011}. 

We fit the two regimes for the relaxation time delimited by an onset temperature for slow dynamics $T^\ast$~\cite{coslovich2007a}. For $T>T^\ast$ an Arrhenius form is used, while for lower temperatures the Vogel-Fulcher-Tammann (VFT) equation is fitted~\cite{vogel1921,fulcher1925,tammann1926}. 
\begin{equation} 
\tau_\alpha^A=
\begin{cases}
\tau_{\infty} \exp{ \left( E_{\infty} / T \right) } & \mbox{ for } T \ge T^\ast,\\
\tau_{\infty}' \exp{ \left( \frac{DT_0}{T-T_{0}}  \right) } & \mbox{ for } T<T^\ast.
\end{cases} 
\label{eqTauAlphaA}
\end{equation}
\noindent We set $T^\ast=1.00$ and fit the Arrhenius equation finding $\tau_{\infty}=0.0693$ and $E_{\infty}=2.91$. For the VFT fit the fragility parameter is found to be $D=7.48$, the VFT temperature is $T_0=0.3250$ and $\tau_{\infty}'$ is set to ensure continuity of the fit at $T^\ast$. 
While it is 
possible to use other fitting forms~\cite{desouza2006,tanaka2005b}, and although VFT may be physically reasonable~\cite{cavagna2009}, there remains no clear consensus as to which form best describes data such as those plotted in Fig. \ref{figAngell}~\cite{hecksler2008}. Nonetheless,  following our previous study~\cite{malins2013jcp}, here we
choose this fitting procedure to reflect the onset of slow dynamics (super-Arrhenius) for $T<1$~\cite{berthier2003,berthier2011} and, over the range of temperatures we consider, find good agreement with the assumption of a crossover to a VFT regime.

\section{Structural analysis}
\label{sectionStructuralAnalysis}

\begin{figure*}
\begin{centering}
\includegraphics[width=12.5cm]{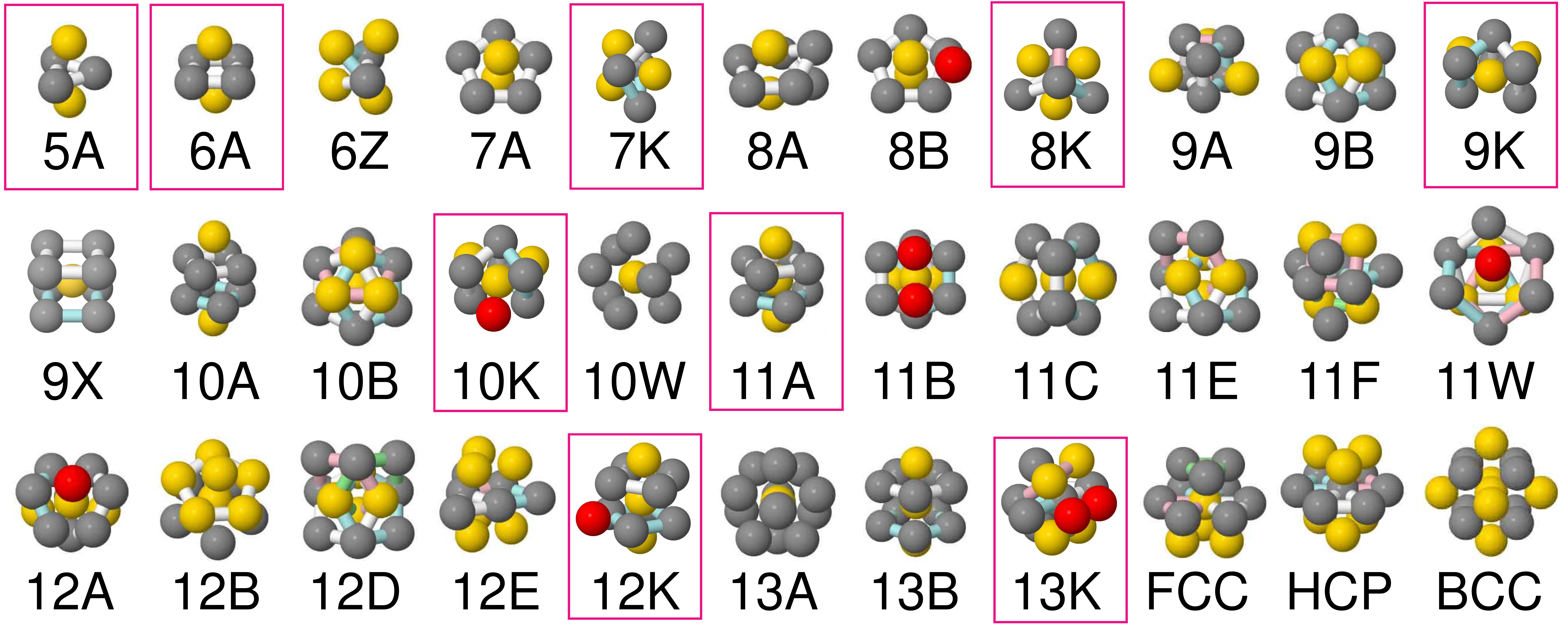}
\par\end{centering} 
\caption{Clusters detected by the topological cluster classification. Highlighted are minimum energy clusters for the Kob-Andersen system. The colours of the particles and the bonds are pertinent to the detection method of the clusters~\cite{malins2013tcc,malinsThesis}.
\label{figTCC}}
\end{figure*}

\begin{figure*}
\begin{centering}
\includegraphics[width=14cm]{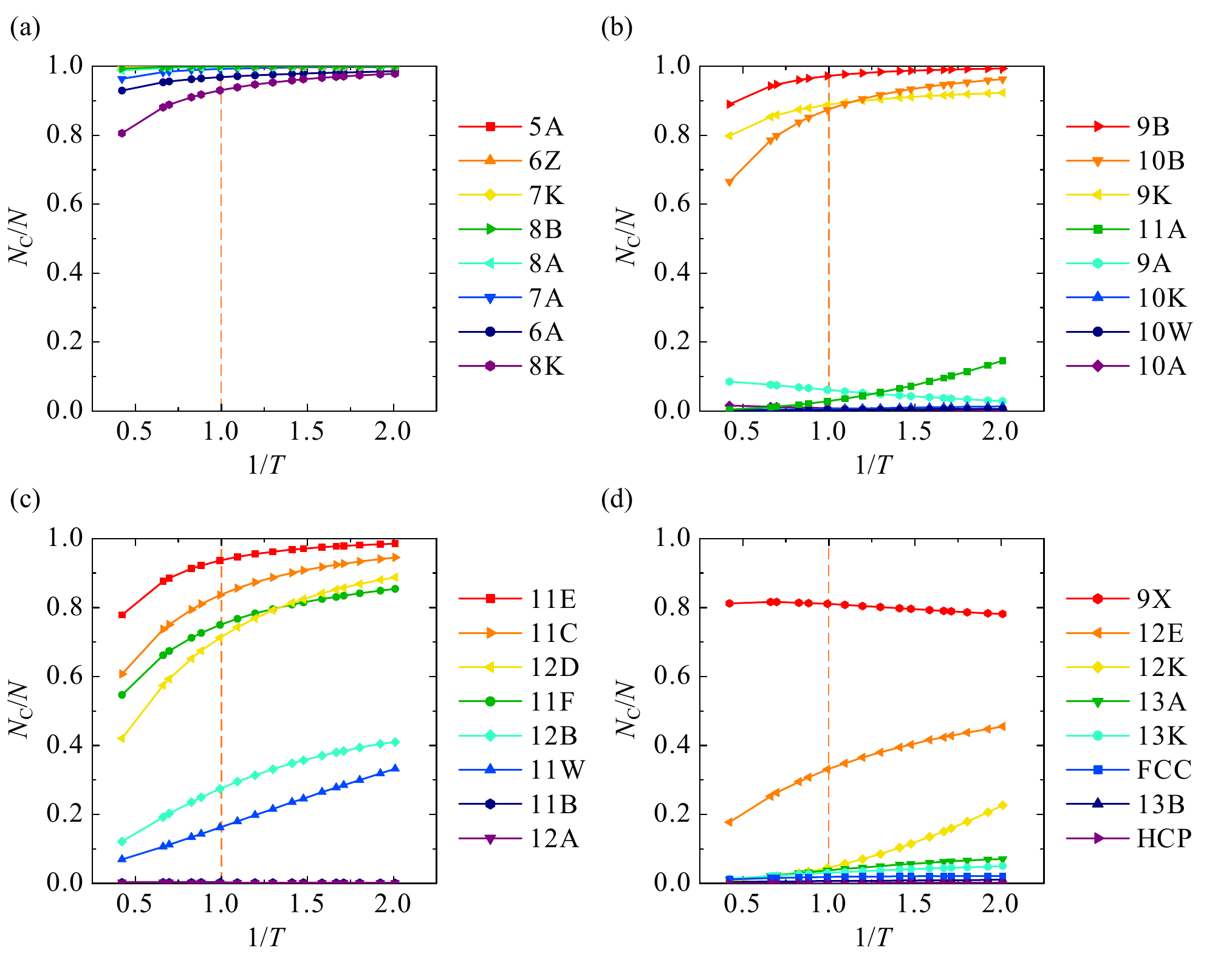}
\par\end{centering}
 \caption{The fraction of particles participating in each cluster type for the KA mixture. The dotted orange lines mark the onset temperature of slow dynamics $T^\ast$. (a) Clusters 5A to 8K, (b) 9A to 11A, (c) 11B to 12D, (d) 12E to 13K and the crystal clusters. 
\label{figGKANcN}}
\end{figure*}

\subsection{Fraction of particles participating within clusters}
\label{sectionFraction}

We analyse how the particles in the supercooled liquids are structured using the topological cluster classification algorithm. This algorithm identifies a number of local structures as shown in Fig. \ref{figTCC}, including those which are the minimum energy clusters for $m=5$ to $13$ KA particles in isolation. The first stage of the TCC algorithm is to identify the bonds between neighbouring particles. The bonds are detected using a modified Voronoi method with a maximum bond length cut-off of $r_\mathrm{c}=2.0$ for all types of interaction (AA, AB and BB)~\cite{williams2007,malins2013tcc,malinsThesis}. A parameter which controls identification of four- as opposed to three-membered rings $f_\mathrm{c}$ is set to unity thus yielding the direct neighbours of the standard Voronoi method~\citep{meijering1953,brostow1978,medvedev1986,malins2013tcc,malinsThesis}.

In Fig. \ref{figGKANcN} we plot the fraction of particles detected within each type of cluster, $N_\mathrm{C}/N$. The onset temperature for slow dynamics is indicated by the orange dotted line on each of the plots. It is clear from these order parameters that the liquid sees continuous changes in local structure as it is cooled. The majority of clusters see an increase in their numbers upon cooling. All of the particles are identified within certain simple structures such as the 5A triangular bipyramid irrespective of temperature, while other more complicated clusters, such are 10K, 11B and HCP, are almost never seen.

Given the variety and range of structural changes that occur, it is not clear \textit{a priori} which of the structures, if any, are important for the formation of dynamic heterogeneities on cooling. It should not be assumed that just because structural changes occur within the supercooled regime that they are responsible for the formation of dynamic heterogeneities, as structural changes also occur in the Arrhenius regime which is not characterised by glassy behaviour. The numbers of particles within 11A, 11W, 12B, 12E and 12K clusters show the largest relative increases in the super-Arrhenius regime. Moreover, the rate of increase in these clusters grows upon further supercooling.

We also find clusters in which almost all particles participate, while other clusters are only seen in trace quantities. In general the trend for clusters where $N_\mathrm{C}/N$ changes significantly on cooling is for it to increase monotonically. However this is not always the case, as seen for the numbers of particles within 9A and 9X clusters which decrease on cooling. The question remains as to how to determine the contribution of the different structures to the glassy behaviour and dynamical heterogeneities of the supercooled liquid.

\subsection{Cluster lifetime distributions}
\label{sectionClusterLifetimeDistributions}

In order to identify which clusters might be relevant to the slow dynamics, we employ the \emph{dynamic} topological cluster classification algorithm~\cite{malins2013jcp,malins2013tcc} to measure the lifetimes of the different TCC clusters at the lowest temperature state point. A lifetime $\tau_{\ell}$ is assigned to each ``instance'' of a cluster, where an instance is defined by the unique indices of the particles within the cluster and the type of TCC cluster. Each instance cluster occurs between two frames in the trajectory and the lifetime is the time difference between these frames. Any periods where the instance is not detected by the TCC algorithm are shorter than $\tau_\alpha^A$ in length, and no subset of the particles becomes un-bonded from the others during the lifetime of the instance.

The measurement of lifetimes for all the instances of clusters in these $N=10976$ simulations is intensive in terms of the quantity of memory required to store the instances, and the number of searches through the memory required by the algorithm each time an instance of a cluster is found to see if it existed earlier in the trajectory. Therefore we do not measure lifetimes for the clusters where $N_\mathrm{C}/N>0.8$, since the vast majority of particles are found within such clusters and it is not immediately clear how dynamic heterogeneities could be related to structures that are pervasive throughout the whole liquid.

\begin{figure}
\begin{centering}
\includegraphics[width=8cm]{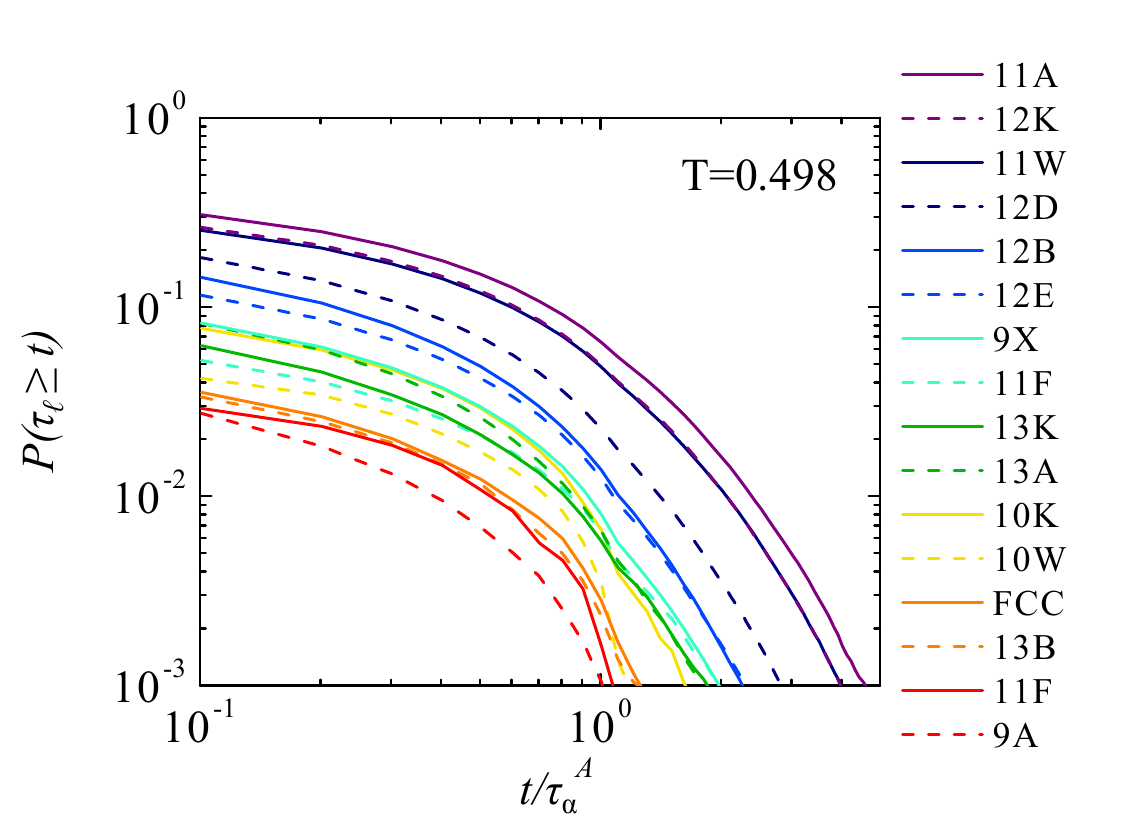}
\par\end{centering}
 \caption[Lifetime autocorrelation functions for clusters in supercooled liquids.]{Lifetime autocorrelation functions for the clusters $P(\tau_{\ell} \ge t)$ for the lowest temperature state point $T=0.498$. 
\label{figGPlt}}
\end{figure}

In Fig. \ref{figGPlt} we plot the lifetime autocorrelation function $P(\tau_{\ell} \ge t)$ for $T=0.498$. Figure \ref{figGPlt} clearly shows that the most persistent or the longest lived of the different types of clusters are the 11A bicapped square antiprisms. All other clusters display lifetime autocorrelation functions that decay more quickly than 11A. The long-time tail of the 11A autocorrelation function indicates that some of these clusters preserve their local structure on timescales far longer than $\tau_{\alpha}^A$. As we shall see below, this effect is enhanced when the 11A group into domains.

The two next-slowest decaying clusters are 11W and 12K. 12K is a KA minimum energy cluster, formed by bonding an additional particle to 11A. There is a high degree of overlap between the 11W clusters and the 11A clusters as their bonding is similar, and only small fluctuations are required for an 11A to be reclassified as an 11W. However the faster-decaying clusters also contain KA minimum energy clusters, for example 13K and 10K. Moreover the lifetimes of all cluster types hold no simple relationship to their size and frequency of occurrence. For example the $n=11$ particle 11F cluster is much more numerous than 11A, yet displays far quicker decay of $P(\tau_\ell \ge t)$. There is also no monotonic trend in the lifetime of the ground state clusters with the cluster size, as the 10K and 13K decay faster than the 12K and 11A. These results clearly demonstrate that the average lifetime of each type of cluster is a property of the local ordering of the particles rather than the size of the cluster or its pervasiveness.  

The fast initial drops of $P(\tau_\ell \ge t)$ reflect the existence of large numbers of clusters with lifetimes $\tau_\ell \ll \tau_\alpha^A$. The lifetimes of these clusters are comparable to the timescale for beta-relaxation where the particles fluctuate within their cage of neighbours. It could be argued that these clusters arise spuriously due to the microscopic fluctuations within the cage, and that the short-lived clusters are not representative of the actual liquid structure. However almost no 11A are found at higher temperatures, cf. Fig. \ref{figGKANcN}(b), where microscopic fluctuations in the beta-regime also occur. We have not yet found a way to distinguish between the short and long-lived 11A structurally, so we conclude that the measured distribution of 11A lifetimes, which includes short-lived clusters, is representative of the true lifetime distribution. However, given the structural similarity of 11A and 11W, small fluctuations leading to reclassification could contribute to the drop at short times. We will see below that as 11A overlap, one particle may be a member of multiple clusters and that the majority of particles found in short-lived 11A also participate in longer-lived 11A. In other words short- and long-lived 11A mainly lie in the same regions of the liquid. 

Another interpretation for the initial drop in $P(\tau_\ell>t)$ of the clusters is that our intuition that there is constant local structure in the beta-regime, which then relaxes on the timescale of the alpha-regime, is incorrect. It may be the case that microscopic ballistic and ``cage-rattling'' motions are enough to reorder local structures without relying on the ``cage-hopping'' motions of the alpha-regime. It has been seen in previous studies that deeply supercooled liquids can crystallise on a timescale before the diffusive range of the mean squared displacement is reached, and occurs with most particles moving by less than one diameter~\cite{zaccarelli2009,saikaVoivod2009,sanz2011,taffs2013}. Those results demonstrate that significant changes in local structure (i.e. liquid-like to crystalline) are possible with only small movements of the particles, which could explain the initial drops of $P(\tau_\ell \ge t)$ that occur on a timescale $\simeq0.1\tau_\alpha^A$.

\subsection{Composition and dynamics of particles in long-lived clusters}
\label{sectionClusterComposition}

The structural analysis above was performed by treating all particles identically with the TCC algorithm. Here we examine the composition of the 11A bicapped square antiprisms in terms of $A$- and $B$-species. The 11A cluster consists of a central particle surrounded by 10 outer (or shell) particles. We find that the central particle is a $B$-specie in $>99\%$ of all instances of 11A clusters. This is a different composition than the crystal structures found to be low-lying energy minima for the KA mixture~\cite{middleton2001}, and may be related to the stoichiometry of the system. Thus the 11A we find are unrelated to any underlying crystal.

\begin{figure}
\begin{centering}
\includegraphics[width=7cm]{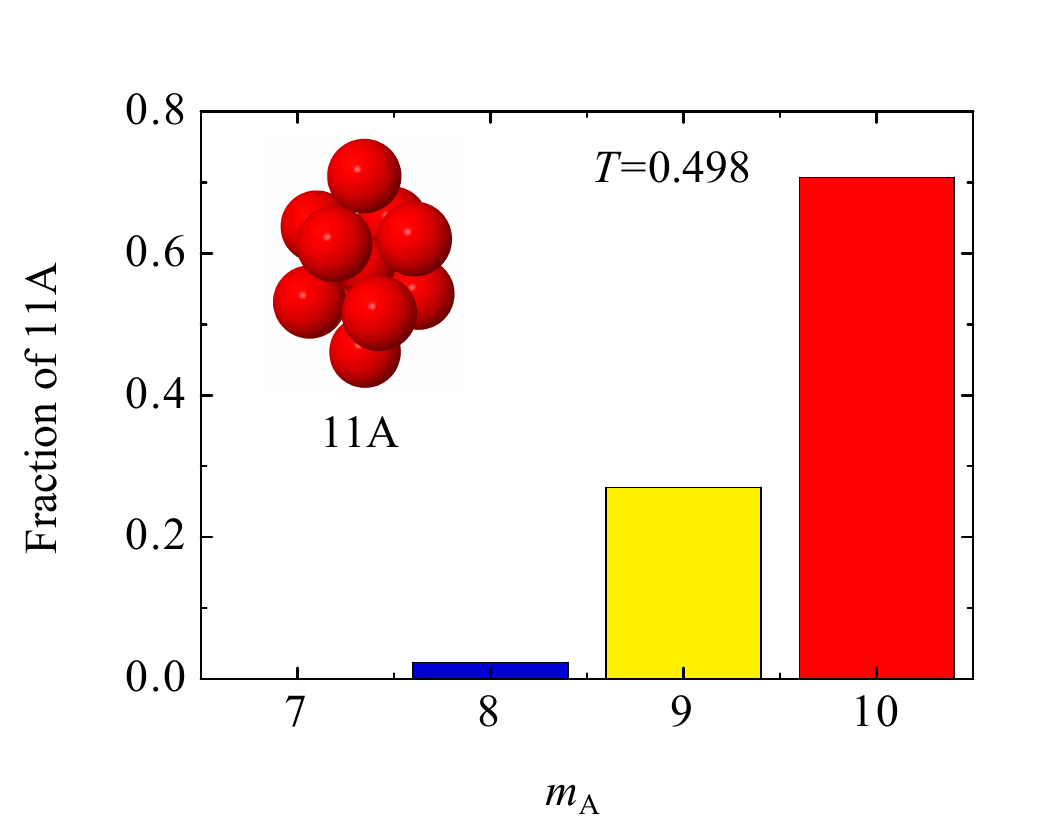}
\par\end{centering}
 \caption{Composition of the 11A clusters at $T=0.498$. Almost all 11A have a small $B$-particle at the centre. $m_A$ is the number of $A$-species in the shell of the cluster. The height of the bars show the relative proportions that each of the compositions occurs. 
\label{figGComp}}
\end{figure}

In Fig. \ref{figGComp} we plot the compositions of the shell particles of the 11A clusters. The majority of 11A clusters have $m_A=10$ $A$-species in the shell of the cluster. We note that this arrangement maximises the number of $AB$ bonds for the central $B$-particle, which is energetically favourable for the central $B$-particle with the KA Lennard-Jones interactions.

\begin{figure*}
\begin{centering}
\includegraphics[width=12cm]{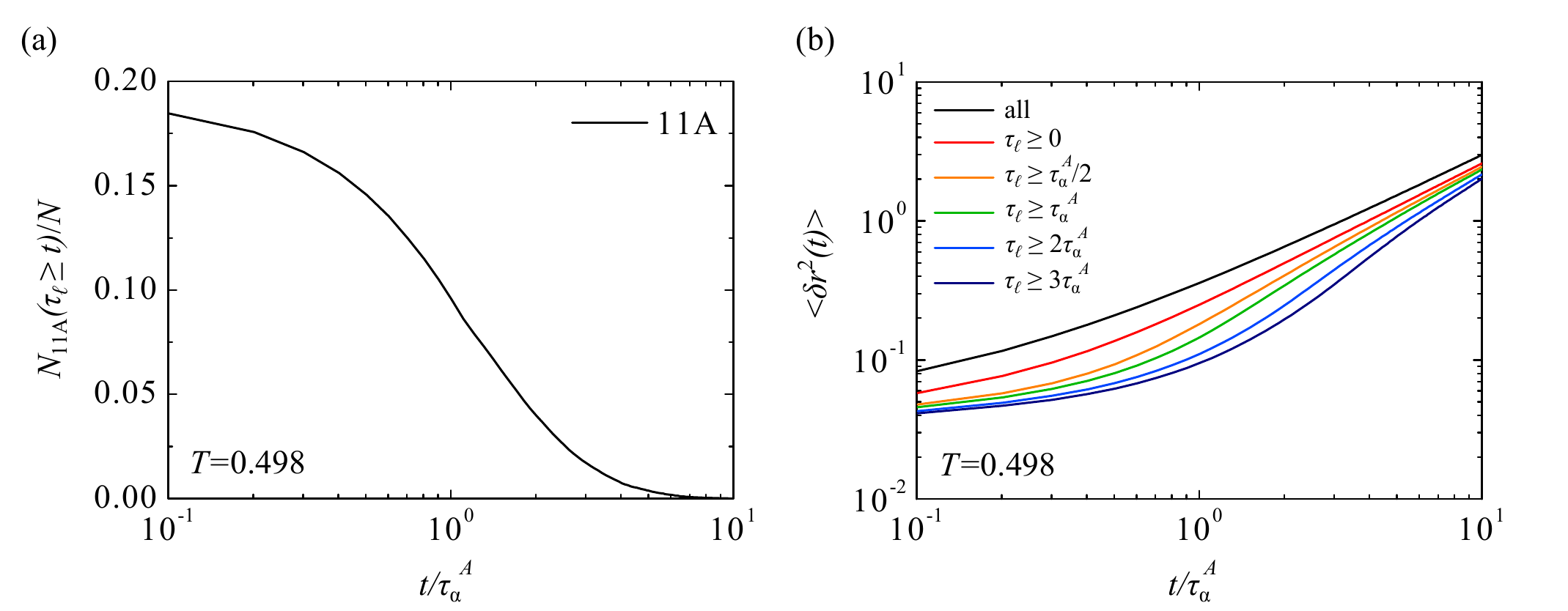}
\par\end{centering}
 
\caption[Dynamics of particles in 11A clusters.]{Dynamics of the particles within 11A clusters in the KA mixture. (a) The fraction of particles participating in 11A clusters with lifetime $\tau_\ell>t$. $N_\mathrm{11A}(\tau_\ell\ge0)/N=0.24$.  (b) The mean squared displacement of particles identified initially within 11A polyhedra of various lifetimes. 
\label{figG11ADyn}}
\end{figure*}

In Fig. \ref{figG11ADyn} we examine how the dynamics of the 11A clusters translates into the dynamics of individual particles. We show in Fig. \ref{figG11ADyn}(a) that the number of particles within 11A clusters as a function of the cluster lifetime. Although there is a fast initial drop in the lifetime autocorrelation function of 11A clusters on the beta-relaxation timescale (Fig. \ref{figGPlt}, solid purple line), as the 11A overlap there remains a significant fraction of the particles these clusters with lifetimes comparable to the dynamic heterogeneities $\approx \tau_\alpha^A$. The difference between $N_\mathrm{11A}(\tau_\ell \ge 0.1\tau_\alpha^A)/N$ and $N_\mathrm{11A}(\tau_\ell \ge 0)/N$ indicates that only $6\%$ of the particles are members of 11A with $\tau_\ell<0.1\tau_\alpha^A$ and \textit{not} a member of an 11A with a longer lifetime as well.

Figure \ref{figG11ADyn}(b) shows the mean squared displacement (MSD) of the particles identified initially within 11A clusters (coloured lines) and compares this to the system-wide MSD (black line). The MSD is defined as the ensemble average $\langle \delta r^2(t) \rangle=\langle |\mathbf{r}_i(t+t_0)-\mathbf{r}_i(t_0)|^2 \rangle$ for the subset of particles of interest (indexed by $i$). All of the particles within 11A relax more slowly than the system-wide average (black line), and the time they take to attain diffusive motion increases as the lifetime of the 11A in which they participate in at $t_0$ increases. In other words the longer the lifetime of the 11A cluster, the slower the particles become. Since some 11A last for very long times [Fig. \ref{figGPlt}], it is expected that these particles may exhibit very low mobilities as they maintain some of their nearest neighbours throughout (e.g. the central particle in an 11A cluster will always have the same shell particles as its nearest neighbours). For the longest lived 11A (blue lines) there appears to be a super-diffusive regime after the initial sub-diffusive regime, indicating that the particles in these clusters may be hopping out of their cage of neighbours as the 11A structure relaxes.

\subsection{Analysis of structured domains}
\label{sectionDomainAnalysis}

\begin{figure*}
\begin{centering}
\includegraphics[width=12cm]{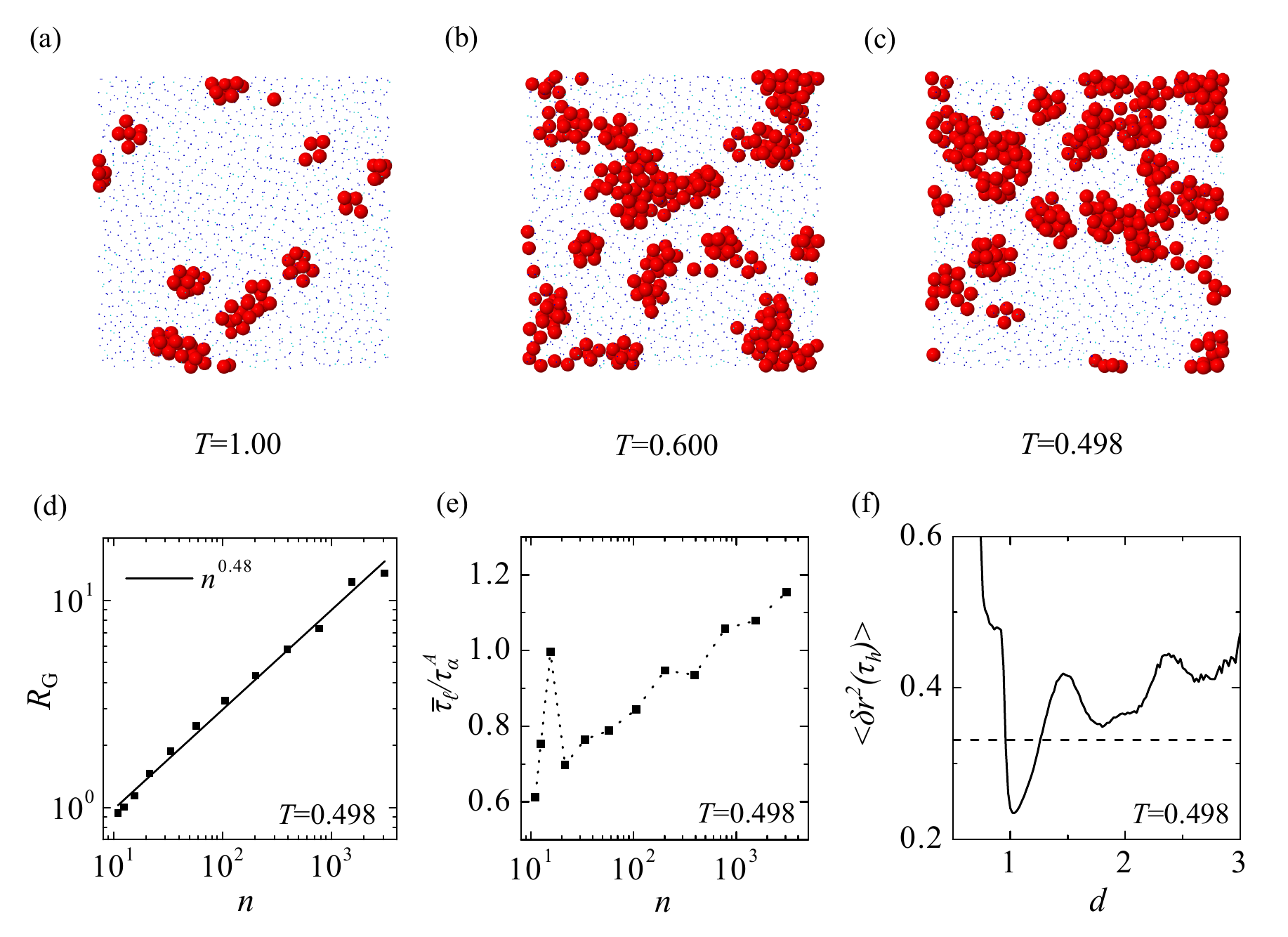}
\par\end{centering}
 
\caption{Analysis of the domains of 11A clusters. (a)-(c) Domains form on cooling from high to low temperature (slices through 3D simulation box). Particles in 11A clusters are shown full size in red, other particles are blue dots. (d) The radius of gyration $R_\mathrm{G}$ of the domains versus the number of particles in the domain $n$ for $T=0.498$. $R_\mathrm{G}$ is well fitted by $n^{0.48}$ indicating the domains have a fractal dimension $d_\mathrm{f} \simeq 2$. (e) The mean lifetime of 11A clusters $\bar{\tau}_\ell$ versus the domain size $n$. (f) 11A domains affect the motion of neighbouring particles. The MSD $\langle \delta r^2(\tau_h) \rangle$ of non-11A particles as a function of distance from 11A domains $d$ (solid line).  The dotted line is the MSD over $\tau_h$ of all particles not in 11A clusters and independent of the distance from an 11A domain ($d$).
\label{figGKAFrac}}
\end{figure*}

On cooling, the number of 11A clusters in the KA mixture increases. At high temperatures the clusters are generally isolated from one another [Fig. \ref{figGKAFrac}(a)]. As the temperature is lowered and the number of clusters increases, domains of clustered particles form. We now analyse the character of these domains of clustered particles and determine the effect the domains have on individual particle dynamics.

The domains of 11A that form on cooling in the KA mixture are shown in Fig. \ref{figGKAFrac}(a)-(c). At high temperature 11A are predominantly isolated [Fig. \ref{figGKAFrac}(a)]. Upon cooling, the 11A overlap and join together [Fig. \ref{figGKAFrac}(b)] to form (transient) networks at low temperatures [Fig. \ref{figGKAFrac}(c)]. 

In order to investigate the structure of the domains, we calculate their radius of gyration,
\begin{equation}
R_\mathrm{G}=\frac{1}{2n^2} \sum_{i,j}(\mathbf{r}_i-\mathbf{r}_j)^2, 
\label{eqRG}
\end{equation}
\noindent where $n$ is the number of particles in the domain and the double sum extends over all pairs of particles in the domain. For configurations where the domains of 11A percolate throughout the simulation box $R_\mathrm{G}$ cannot be defined. These configurations are rare at $T=0.498$ and are excluded from the analysis. The consequences of percolating 11A domains are discussed further below.

The radius of gyration shows a power law growth in the size of the domain $n$ with an exponent of $0.48$ [Fig. \ref{figGKAFrac}(d)]. In other words the domains have a fractal dimension $d_\mathrm{f} \simeq 2$, indicating that they are not space-filling. The individual 11A have enhanced stability as the size of the 11A domains grow [Fig. \ref{figGKAFrac}(e)]. The time $\bar{\tau}_\ell$ is the mean lifetime of 11A clusters constituting a domain of size $n$ in a configuration. For $T=0.498$ the general trend is that the mean 11A lifetime increases with the size of the 11A domains.  Fig. \ref{figGKAFrac}(e) indicates that there is particularly stable arrangement of two overlapping 11A clusters ($n\simeq16$) relative to domains from similar size. The mean lifetime $\bar{\tau}_\ell$ doubles between isolated 11A ($n=11$) and extended domains ($n \approx 1000$).

We now consider the effect the domains have on the remainder of the system. In Fig. \ref{figGKAFrac}(f), we plot the MSD of the particles not in 11A domains $\langle \delta r^2(\tau_h) \rangle$ against the distance $d$ from the nearest 11A particle at time $t=t_0$. The time $\tau_h\simeq\tau_\alpha^A$ is the time of the maximum in the dynamic susceptibility $\chi_4(t)$ defined below~\cite{lacevic2003}. For distances $d<0.96$ the non-11A particles are mainly B-species due to the nonadditive nature of the KA Lennard-Jones interactions. These particles are more mobile than the majority A-species, independent of $d$, due to their smaller size. The number of particles with $d<0.96$ of a 11A domain is around $10\%$ of the system, i.e. half of all the B-species. 

Moving further away from the 11A domains, there is then a region of particles with $d\simeq1$ with reduced mobility compared to the average for non-11A particles [dotted line in Fig. \ref{figGKAFrac}(f)]. Around $37\%$ of the particles are found in this region. Subsequent neighbours of the 11A domains for $d>1.26$ have increased mobility relative to the average. Therefore, excluding the minority $B$-species, the first nearest neighbours of the domains have suppressed mobility compared to the average, indicating coupling between the structured domains of 11A clusters and the dynamics of the neighbouring particles. Correspondingly the second and third shells of neighbours to the 11A domains have higher mobility compared to the average, indicating a hierarchy of spatial dynamics related to the domains of 11A clusters.

\section{Correlation lengths}
\label{sectionCorrelationLengths}

\begin{figure}
\begin{centering}
\includegraphics[width=6cm]{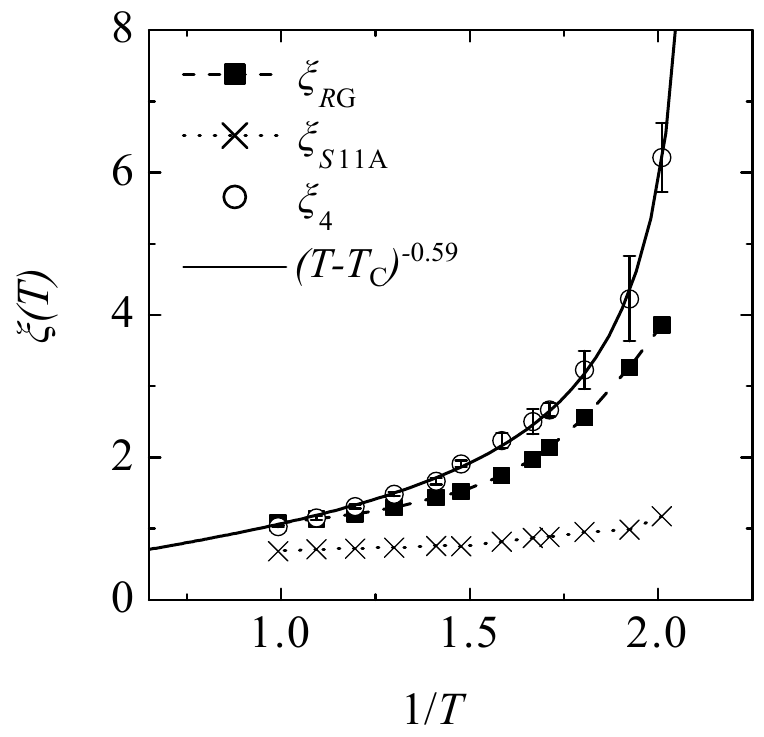}
\par\end{centering}
\caption{Comparison of static and dynamic correlation lengths. Structural $\xi_{RG}$ (squares) and $\xi_{S\mathrm{C}}$ (crosses), and dynamical  $\xi_{4}$ (circles) correlation lengths. $\xi_{4}$ is fitted with a power law (solid line), which diverges at a value $T_\mathrm{C}=0.47$.
\label{figGLengths}}
\end{figure}

\subsection{Dynamic correlation lengths}

Finally we consider whether the structured domains of particles are related to the increasing dynamic correlation lengths in supercooled liquids. In order to do this, we calculate the dynamic correlation length $\xi_{4}$, following La\v{c}evi\'{c} \textit{et al.}~\cite{lacevic2003}. Details of this procedure can be found elsewhere~\cite{lacevic2003,malins2013jcp,flenner2009}. $\xi_{4}$ has been previously calculated for the Kob-Andersen model  and our values correspond closely to those in the literature~\cite{flenner2009}. The dynamical correlation length $\xi_4$ is obtained by analogy to critical phenomena~\cite{lacevic2003}. A (four-point) dynamical susceptibility is calculated as
\begin{equation}
\chi_4(t)=\frac{V}{N^2 k_B T}[\langle Q(t)^2\rangle -\langle Q(t)\rangle^2],
\label{eqChi4}
\end{equation}
\noindent where 
\begin{equation}
Q(t)=\frac{1}{N} \sum_{j=1}^N \sum_{l=1}^N w(|\textbf{r}_j(t+t_0)-\textbf{r}_l(t_0)|).
\label{eqQ}
\end{equation}
\noindent The overlap function $w(|\textbf{r}_j(t+t_0)-\textbf{r}_l(t_0)|)$ is defined to be unity if $|\textbf{r}_j(t+t_0)-\textbf{r}_l(t_0)|\le a$, 0 otherwise, where $a=0.3$. The dynamic susceptibility $\chi_4(t)$ exhibits a peak at $t=\tau_h$, which corresponds to the timescale of maximal correlation in the dynamics of the particles.
We then construct the four-point dynamic structure factor $S_4(\textbf{k},t)$:
\begin{eqnarray*}
S_4(\textbf{k},t)&=&\frac{1}{N\rho} \langle \sum_{jl} \exp[-i \textbf{k} \cdot \textbf{r}_l(t_0)]w(|\textbf{r}_j(t+t_0)-\textbf{r}_l(t_0)|)\\
&\times& \sum_{mn} \exp[i \textbf{k} \cdot \textbf{r}_n(t_0)]w(|\textbf{r}_m(t+t_0)-\textbf{r}_n(t_0)|) \rangle,
\end{eqnarray*}
\noindent where $j$, $l$, $m$, $n$ are particle indices and $\mathbf{k}$ is the wavevector. For time $\tau_h$, the angularly averaged version is $S_4(k,\tau_h)$. The dynamic correlation length $\xi_4$ is then calculated by fitting the Ornstein-Zernike (OZ) function to $S_4(k,\tau_h)$, as if the system were exhibiting critical-like spatio-temporal density fluctuations, 
\begin{equation}
S_4(k,\tau_h)=\frac{S_4(0,\tau_h)}{(1+(k\xi_4(\tau_h))^2)},
\label{eqOZ}
\end{equation}
\noindent to $S_4(k,\tau_h)$ for $k<2$~\cite{lacevic2003}. The resulting $\xi_4$ are plotted in Fig. \ref{figGLengths}. 

We carried out an unconstrained fit to the $\xi_4$ data, according to $\xi_4(T)=\xi_4^0(T-T_\mathrm{C})^{-\nu}$. The line is plotted in Fig. \ref{figGLengths}. We find the ``critical exponent'' is $\nu =0.588 \pm 0.02$, the ``critical temperature'' is $T_\mathrm{C}=0.471 \pm 0.002$ and the prefactor is $\xi_4^0 =0.59 \pm 0.02$. Under the caveat that obtaining $\xi_4$ from fitting $S_4$ in limited size simulations is notoriously problematic~\cite{flenner2007,flenner2009} and thus any numerical values should be treated with caution, we observe that the value of $T_\mathrm{C}$ is not hugely different to the glass transition temperature found by fitting
Mode-Coupling theory to this system, around 0.435~\cite{kob1995a,kob1995b}. We also note that $\nu =0.588$ lies between mean field ($\nu =0.5$) and 3D Ising ($\nu =0.63$) criticality.
Here it is worth noting that Onuki and his coworkers pointed out that 
the dynamical correlation length estimated from four-point density correlations may seriously be affected by 
thermal low-frequency vibration modes, which may lead to a strong system-size dependence~\cite{shiba2012}. 
They also showed that a bond-breakage correlation length is free from these vibrational modes, and thus 
a suitable measure of dynamical coherence.

\subsection{Static correlation lengths}

We consider two static correlation lengths for the domains of particles in 11A clusters. The first method allows for direct comparison with the dynamic lengthscale $\xi_4$. We define a structure factor restricted to the particles identified within 11A:
\begin{equation}
S_\mathrm{11A}(\textbf{k})=\frac{1}{N\rho} \langle \sum_{j=1}^{N_\mathrm{11A}} \sum_{l=1}^{N_\mathrm{11A}} \exp[-i \textbf{k} \cdot \textbf{r}_j(t_0)]\exp[i \textbf{k} \cdot \textbf{r}_l(t_0)] \rangle,
\label{eqS11A}
\end{equation} 
\noindent where $N_\mathrm{11A}$ is the number of particles in 11A clusters. We then fit the Ornstein-Zernike equation (Eq. \ref{eqOZ}) to the low-$k$ behaviour of the angularly-averaged $S_\mathrm{11A}(k)$ in order to extract a structural correlation length $\xi_{S\mathrm{11A}}$. This is plotted in Fig. \ref{figGLengths}. This procedure is akin to the calculation of the dynamic lengthscale $\xi_4$: first a structure factor is calculated from a selected fraction of the particles (either immobile or structured), and the Ornstein-Zernike expression used to extract a correlation length. 

The second lengthscale we consider for the structured particles is derived from the radius of gyration of the domains of clusters. We define
\begin{equation}
\xi_{R\mathrm{G}}=  R_\mathrm{G}^{\mathrm{11A}} (\langle n \rangle / m)^{1/d_\mathrm{f}},
\label{eqxiRG}
\end{equation} 
where $R_\mathrm{G}^{\mathrm{11A}}$ is the radius of gyration of a single cluster, $\langle n \rangle$ is the ensemble average of the domain size, $m$ is the number of particles in the cluster, and $1/d_\mathrm{f}$ is the exponent of the power law fitted to $R_\mathrm{G}$ versus $n$. This correlation length does not probe the correlations between the domains, as per $\xi_{S\mathrm{11A}}$, rather it characterises the growth in size of the domains on cooling until a percolation transition is reached.

The temperature behaviour of the different correlation lengths is shown in Fig. \ref{figGLengths}. All three correlation lengths increase on cooling, however the manner in which each of the lengths increases is quite different. The main result is that the growth in the dynamic correlation length $\xi_4$ is not matched by the growth in the structural correlation length $\xi_{S\mathrm{11A}}$. Thus we do not find one-to-one correspondence between the behaviour of structural and dynamic correlation lengths.

\subsection{Discussion}
\label{sec:discussion}

The fact that different behaviour between the dynamic and static lengths is found is in agreement with some recent studies on 2D and 3D systems~\cite{hocky2012,dunleavy2012,charbonneau2012,charbonneau2013,charbonneau2013a}, however we note that a one-to-one correspondence in the growth in a lengthscale relating to static crystalline order and the dynamic correlation length for polydisperse quasi-hard sphere systems has been found~\cite{shintani2006,kawasaki2010,tanaka2010,sausset2010}.

The structural order we find is distinct from the crystalline structure and is thought to frustrate crystallisation. We note that in the studies on 3D systems that have found one-to-one correspondence between the structural and dynamical lengthscales~\cite{kawasaki2010,tanaka2010}, the relative increase in the static lengthscales going from state points in the Arrhenius regime to into the supercooled regime is no more than a factor of $2$. Increases of comparable magnitude have been found in studies using point-to-set measures~\cite{hocky2012}, and more indirect methods~\cite{mosayebi2010,mosayebi2012} rather than focussing on explicit structures. 
In Fig. \ref{figGLengths} our static length $\xi_{S\mathrm{11A}}$ shows an increase of the same order between Arrhenius and supercooled state points, however the increase in $\xi_4$ is relatively much greater (factor of $\approx 6$ or more between high and low $T$). We note a recent 2D study that suggests one-to-one correspondence in lengthscales for crystalline order and dynamic heterogeneities on quasi-hard spheres breaks down with the addition of attractions between the particles~\cite{xu2012}. However, we also note that 3D polydisperse hard-sphere and Lennard-Jones systems have the same link between the correlation length of crystal-like order and slow dynamics, albeit over a limited range of correlation length~\cite{tanaka2010}. These points need to be clarified in the future.  

The structured domains form rarefied networks with $d_\mathrm{f} \simeq 2$. These networks do not fill space, whereas it is thought that any critical-like nature of the dynamic heterogeneities would imply that the domains have $d_\mathrm{f} \simeq 2.5$~\cite{onuki}. Therefore geometrically it appears that the domains of slow and structured particles may be different, at least at the state points we have accessed. Moreover the number of particles in the domains is unlikely to coincide across a range of temperatures, as the $N_\mathrm{11A}$ for the structured particles increase monotonically on approaching the glass transition, while the number of particles selected by $Q(t)$ is relatively constant as a function of temperature.

Here we note that Mosayebi et al.~\cite{mosayebi2010,mosayebi2012} found behaviour consistent with critical-like static correlations in the same system as ours, which implies $d_f \sim 2.5$. This seems to suggest that 11A together with other clusters or kinetically pinned 
particles may correspond to their static structures with solid-like nature. Now the temperature range over which we were able to equilibrate our simulations is less than that for which
Mosayebi \emph{et al.}~\cite{mosayebi2010,mosayebi2012} present data. Indeed over our range of temperature, the relative increase in static lengthscale they measure is consistent with ours. However, since our $d_f \sim 2.0$, our findings are not consistent with Ising-like criticality, 
although critical-like behaviour at lower temperatures than we have been able to access cannot be ruled out. We also note that the structural motifs investigated here are of intrinsically of discrete nature, whereas the structural measures showing critical-like behaviours ($d_f \sim 2.5$), such as bond orientational order \cite{tanaka2010,tanaka2012} and static structures \cite{mosayebi2010,mosayebi2012} have a continuous nature.

The MSD of the 11A particles indicates that there is not a one-to-one correspondence between the particles selected in $S_4$ and $S_\mathrm{11A}$. The particles selected by $w$ for $S_4$ all have $\delta r^2(\tau_h) \le 0.09$ strictly (by the definition of $w$). However, as can been seen from the red line in Fig. \ref{figG11ADyn}(b), the MSD of the 11A particles over the same timescale is $\langle \delta r^2(\tau_h) \rangle\approx 0.1$. Only the particles in the longest-lived 11A clusters [blue lines in Fig. \ref{figG11ADyn}(b)] have squared displacements over $\tau_h$ comparable to the immobile particles on which $S_4$ is measured. Thus on this basis direct correspondence between $\xi_4$ and $\xi_\mathrm{11A}$ should not necessarily be expected. We note that this might also be related to the effects of low-frequency vibrational modes~\cite{shiba2012}.

Note $\chi_4$ has been shown to exhibit dependence upon system size for $N \lesssim 1000$~\cite{karmakar2009}. We expect that such effects are reasonably small here, and have taken care to only consider temperatures where all our measured lengths are smaller than the system size. While system size effects cannot be ruled out, we do not believe these make a significant impact on the conclusions we draw.

The lengthscale $\xi_{R\mathrm{G}}$ indicates how the size of the domains grows on cooling. In our related study on the Wahnstr\"{o}m binary Lennard-Jones glass former~\cite{malins2013jcp}, where the concentration of clusters was higher, a percolating network of icosahedra was formed. In a study on the Kob-Andersen model considered here~\cite{speck2012}, some of us found evidence for an increase in the population of 11A clusters at lower temperatures than we have been able to access here. Thus if the 11A clusters percolate, this would suggest a diverging structural lengthscale at a temperature higher than either the VFT or MCT temperatures ($\sim0.325$ and $\sim0.435$ respectively). In any case, percolating domains of 11A clusters do not imply structural arrest since each cluster has a finite lifetime. This scenario contrasts with colloidal gels where a percolating network of local structures leads to dynamic arrest~\cite{royall2008}.

We also note that Fig. \ref{figGKAFrac}(f) strongly indicates that the effect of the domains of clustered particles on the surrounding liquid extends around one particle diameter from the domains. This shows that the dynamical effect of the structured particles is hierarchical and not solely limited to the structured domains themselves. However the correlation lengths we measured from a structure factor of the domains and their first nearest neighbours was no greater than the lengthscale $\xi_{S\mathrm{11A}}$ for the domains themselves.

Thus the question remains as to the most appropriate structural (and dynamical) correlation length to explain the viscous slowing down in supercooled liquids, and how order-specific correlation lengths, such as $\xi_{R\mathrm{G}}$ and $\xi_{S\mathrm{11A}}$, are related to order-agnostic structural correlation lengths~\cite{biroli2008,mosayebi2010,sausset2011,dunleavy2012,cammarota2012,mosayebi2012}. We conclude this discussion with the following observations. 

\begin{enumerate}
\item Static correlation lengths have been measured in a variety of systems. Most seem to grow less strongly than dynamic correlation lengths in the regime accessible to computer simulation (and real-space colloid experiments)~\cite{biroli2008,mosayebi2010,sausset2011,dunleavy2012,cammarota2012,mosayebi2012,hocky2012,malins2013jcp}. Those that are observed to grow in a way comparable to the dynamic correlation length $\xi_4$ are often related to crystalline order~\cite{tanaka2010,shintani2006,kawasaki2007}. 
Here we note that a bond orientational order parameter is a continuous variable, whereas structural motifs are discrete by the definition. There might also be some effects of low-frequency vibrational modes on the estimation of the dynamical correlation length in our system~\cite{shiba2012}. These points need further investigation.
\item We note that geometric frustration suggests a term in $R^5$ to be included in a classical nucleation theory like equation, where $R$ is the size of a growing domain of the preferred structure (11A here)~\cite{tarjus2005}. This $R^5$ term strongly suppresses growth of such domains, consistent with ramified structures as we (and others~\cite{coslovich2007a,dzugutov2002}) find. Now the arguments supporting geometric frustration tended to focus on icosahedra, not 11A bicapped square antiprisms as we find here. However, in our study of the Wahnstr\"{o}m model (whose local structural motif is the icosahedron)~\cite{malins2013jcp}, we found very similar behaviour to that reported here. In any case, crystals can be formed of 11A, though not for the 80-20 composition of the KA mixture~\cite{fernandez2003}, and icosahedra (with the inclusion of Frank-Kasper bonds)~\cite{pedersen2010}. Assuming crystallisation is avoided, given the presence of such an $R^5$ term, one might enquire as to why growth of polyhedral domains is expected in the first place. 
\item Apart from $\xi_4$, dynamic correlation lengths have also been measured by other means~\cite{hocky2012,kob2011non}. 
In particular, evidence has been found for non-monotonic behaviour of such dynamic correlation lengths around the Mode-Coupling transition~\cite{kob2011non}. While $\xi_4$ itself does not exhibit non-monotonic behaviour as such, there is evidence that a different scaling is followed for quenches below the Mode-Coupling transition~\cite{flenner2013}. Moreover, that our fit of $\xi_4$ implies divergence at $T=0.47$ suggests that some other scaling might be found at lower temperatures than we have accessed.  Another way to approach the discrepancy in Fig. \ref{figGLengths} is to enquire whether  $\xi_4$ is the ``right'' dynamical correlation length~\cite{harrowell2010}. 
One could even speculate that dynamical correlations are enhanced around the Mode-Coupling transition (as our results imply, along with those of Kob \emph{et al.}~\cite{kob2011non}), and that at least the lengthscale does not grow significantly at lower temperatures. This would rationalise the estimates of dynamic correlation lengths close to the \emph{molecular} glass transition (some 10 orders of magnitude slower in relaxation time than our simulations are able to access) which suggest correlation lengths only of a few molecular diameters~\cite{berthier2005,fragiadakis2011}. 
However, we should note that there is a possible deficiency of the standard dynamical correlation length $\xi_4$~\cite{shiba2012}. We point out that the wavenumber dependence of the transport coefficient, viscosity, is a physically appealing method for estimating the intrinsic dynamical correlation length~\cite{furukawa2009,furukawa2011,furukawa2012}.
\end{enumerate}
 
\section{Summary and conclusions}

We have demonstrated that by studying the lifetimes of different structural orderings within the Kob-Andersen supercooled liquid, the relatively stable orderings of particles can be detected unambiguously. This method alleviates some of the difficulties in identifying structural correlations relevant to glassy behaviour from the temperature dependency of the number of particles participating in clusters. 
The most stable cluster found in the Kob-Andersen supercooled liquid is the 11A bicapped square antiprism. The relaxation of particles within these clusters proceeds more slowly as the lifetime of the cluster increases. This is consistent with previous work based on Voronoi polyhedra~\cite{coslovich2007a}.

The long-lived clusters form rarefied domains on cooling with a fractal dimension $d_\mathrm{f}\simeq2$, i.e. the structured domains are non-space-filling, at least in the regime we have accessed. The lifetime of the 11A clusters increases markedly with the size of the domains of these clusters. The non-11A particles neighbouring these domains have reduced mobility compared to particles further from the domains, suggesting a link between structure and dynamic heterogeneity at this level of detail. In other words, the network of 11A clusters acts to ``pin'' its neighbours. 

We examined the relationship between the structured domains and the dynamic heterogeneities by considering static and dynamic correlation lengths of structured/slow particles. A static correlation length calculated in a like-for-like manner with the dynamic correlation length was found to grow moderately on cooling, however its increase was outmatched by the growth in the dynamic correlation length $\xi_4$. The difference in behaviour of the correlation lengths was rationalised by noting that the structured domains grow in a non-space filling manner, and that the correlation between the structured and slow particles is not perfect. The relationship between the our static and dynamic correlation lengths, and other lengthscales for static order, remains an open question, (see sec. \ref{sec:discussion}).

Finally we consider a possible direction for future study. It has been shown recently that an order parameter associated with the population of 11A clusters can be used to drive a first order transition in an ensemble of trajectories~\cite{speck2012}. The susceptibility of the transition to the field coupled to the structural order parameter was found to be higher than when biasing with a field coupled to the dynamical activity, which is the usual method that the transition is accessed. Furthermore a recent study by Singh \textit{et al.} has shown that ``ultra-stable'' KA glasses prepared by a vapour deposition technique have high numbers of clusters equivalent to 11A polyhedra~\cite{singh2013}. Together these results provide further evidence for a connection between the atomic level structure, most easily accessed with high-order structural correlation functions, and the glass transition. 
The biasing of fields coupled to structural order parameters in trajectory space, and possibly in configuration space as well, thus opens up a new route for the preparation of ultra-stable glassy states~\cite{singh2013} pertinent to temperatures well below the Mode-Coupling temperature. The relaxation times inferred for these states are many orders of magnitude higher than those which can currently be prepared with conventional simulations. Study and characterisation of the properties of these states will shed further light on nature and role of local structure in the glass transition. 
In other words, while Fig. \ref{figGLengths} (like much of the recent literature~\cite{hocky2012,dunleavy2012,malins2013jcp,mosayebi2010,mosayebi2012,kob2011non}) indicates a decoupling between structural and dynamical lengthscales, other static and/or dynamical measures might finally lead to coupling of structure and dynamics. 

\section*{Acknowledgements}

We gratefully acknowledge stimulating discussions with Rob P.  Jack, Thomas Speck and Stephen Williams.
A.M. is funded by EPSRC grant code EP/E501214/1. C.P.R. thanks the
Royal Society for funding. H.T. acknowledges support from a grant-in-aid from the Ministry of Education, Culture, Sports, Science and Technology, Japan and the Aihara Project, the FIRST program from JSPS, initiated by CSTP. This work was carried out using the computational
facilities of the Advanced Computing Research Centre, University of
Bristol.


\end{document}